\documentclass[dvips]{article}

\usepackage{icrc2011}
\usepackage[percent]{overpic}
\usepackage{color}

\title{H.E.S.S. observations of the Large Magellanic Cloud}
\newcommand{\etal}{\MakeLowercase{\textit{et al. }}} % "et al."
\shorttitle{Komin \etal H.E.S.S. observations of the LMC}

\authors{Nu. Komin$^{1}$, A. Djannati-Atai$^{2}$,Y. Gallant$^{3}$, V. Marandon$^{2}$, C.C. Lu$^{4}$, S. Ohm$^{4,5,6}$ and E. de Ona-Wilhelmi$^{4}$
for the H.E.S.S. collaboration}
\afiliations{$^1$LAPP, CNRS/IN2P3, Annecy-le-Vieux, France\\ 
$^2$APC, CNRS/IN2P3, Paris, France\\
$^3$LUPM, CNRS/IN2P3, Montpellier, France\\
$^{4}$Max-Planck-Institut fuer Kernphysik, Heidelberg, Germany\\
$^{5}$School of Physics \& Astronomy, University of Leeds, UK\\
$^{6}$Department of Physics \& Astronomy, University of Leicester, UK
}
\email{nukri.komin@lapp.in2p3.fr}

\abstract{
The Large Magellanic Cloud (LMC) is a satellite galaxy of the Milky Way at a distance of approximately 48 kpc. Despite its distance it harbours several interesting targets for TeV gamma-ray observations. The composite supernova remnant N 157B/PSR J05367-6910 was discovered by H.E.S.S. being an emitter of very high energy (VHE) gamma-rays. It is the most distant pulsar wind nebula ever detected in VHE gamma-rays. Another very exciting target is SN 1987A, the remnant of the most recent supernova explosion that occurred in the neighbourhood of the Milky Way. Models for Cosmic Ray acceleration in this remnant predict gamma-ray emission at a level detectable by H.E.S.S. but this has not been detected so far. Fermi/LAT discovered diffuse high energy (HE) gamma-ray emission from the general direction of the massive star forming region 30 Doradus but no clear evidence for emission from either N 157B or SN 1987A has been published. 

The part of the LMC containing these objects has been observed regularly with the H.E.S.S. telescopes since 2003. With deep observations carried out in 2010 a very good exposure of this part of the sky has been obtained. The current status of the H.E.S.S. LMC observations is reported along with new results on N\,157B and SN\,1987A.
}
\keywords{VHE gamma-rays; HESS; Large Magellanic Cloud
}

\begin{document}
\maketitle

%Begin the section.
\section{Introduction}

The Large Magellanic Cloud (LMC) is a satellite galaxy of the Milky
Way at a distance of 48\,kpc \cite{distance}, with an apparent
extension of about $10^\circ$ and an inclination angle of $31^\circ$
\cite{inclination}. It harbours several interesting sources, known to
be possible gamma-ray emitters.

In the survey of the Milky Way performed with H.E.S.S. \cite{scan} a number of Pulsar Wind Nebulae (PWNe)
has been detected, and a connection between the spin-down luminosity
of the powering pulsar and the gamma-ray flux of its nebula could be
established \cite{PWNconnection}. The pulsar with
the largest known spin-down power, $\dot E = 4.9 \cdot 10^{38}
\mathrm{erg/s}$, is PSR J0537$-$6910. This pulsar is part of the composite supernova remnant N\,157B located in the LMC. 

SN\,1987A is a very young (24 years) supernova remnant, and the
closest event that happened in recent history. Models for Cosmic Ray acceleration predict a growing VHE gamma-ray emission from hadronic interactions, to rise by a factor of about two for the next decades \cite{BK2006,BKV2011}. No gamma-ray emission has been detected so far from
SN\,1987A \cite{HDGS04, Cang07}.

MeV and GeV gamma-ray emission from the direction of the LMC has been detected with the \textit{Fermi}/LAT telescope \cite{Fermi}. The emission is largely extended (about $5^\circ$) and it is coincident with the massive star forming region 30 Doradus. No MeV and GeV gamma-ray emission has been detected from point-like sources at the positions of N\,157B or SN\,1987A.

Despite the distance of the LMC, which requires deep exposure, the
observation of VHE gamma-ray sources in the LMC has several advantages:
The distance as an important ingredient in modelling and
interpretation is rather precisely known, and the angular separation
of the sources is small, allowing the monitoring of several objects
in the same camera field of view.

 \begin{figure}[!h]
%%  \vspace{5mm}
  \centering
\begin{overpic}[width=0.5\textwidth]{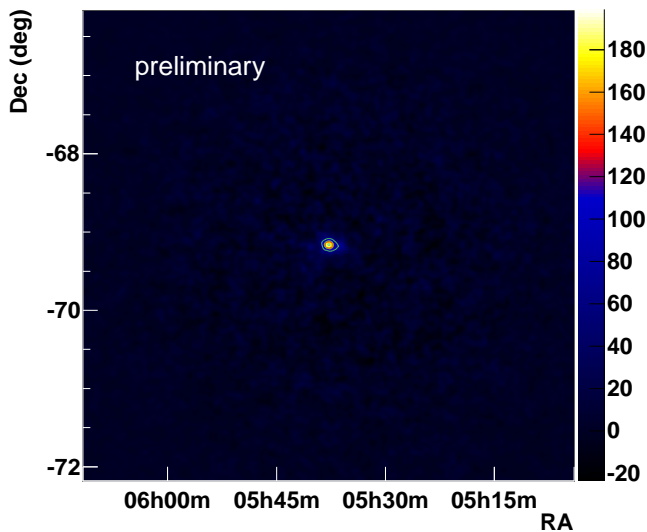}
\put(20,75){\bf \sf {\color{white} preliminary}}
\end{overpic}
  \caption{HESS excess map of the LMC. The excess is smoothed with the point spread function (68\% containment radius of $0.06^{\circ}$).The overlaid contours are obtained from a sky maps oversampled with a radius of $0.06^{\circ}$ and denote 6, 10 and 14\,$\sigma$ significance. }
  \label{fig:skymap}
 \end{figure}

\section{H.E.S.S. Observations}

The High Energy Stereoscopic System (H.E.S.S.) is a system of four
Imaging Cherenkov Telescopes. It can reconstruct the photon's arrival
direction with an angular resolution of $\approx 0.1^\circ$ and
its energy with a resolution of $\approx 20\%$. Its field of view ($5^\circ$) is ideal for the observation of extended
sources and surveys. The sensitivity of H.E.S.S. allows a 5$\sigma$
detection of a source with flux of about 1\% of the flux of the Crab
nebula within 50\,h \cite{Hinton}. Located in the southern hemisphere, H.E.S.S. 
can observe the LMC (Declination $-69^\circ$)
with relatively large zenith angles (around $45^\circ$) resulting in an elevated energy threshold of about 500\,GeV.

The region of the LMC containing N\,157B, SN\,1987A and 30 Doradus has been
observed with a H.E.S.S. on a yearly basis since 2003, with very deep exposure added in 2010. The current exposure is 90.4\,h of live time. Figure~\ref{fig:skymap} shows an excess sky map obtained from these observations. The significance levels were calculated from sky maps which were oversampled with a radius of $0.06^{\circ}$, adapted to the search for point-like sources. In this sky map gamma-ray emission exceeding $14\,\sigma$ statistical significance can be seen.

\section{The Pulsar Wind Nebula N\,157B}

 \begin{figure}[!h]
%%  \vspace{5mm}
  \centering
\begin{overpic}[width=0.5\textwidth]{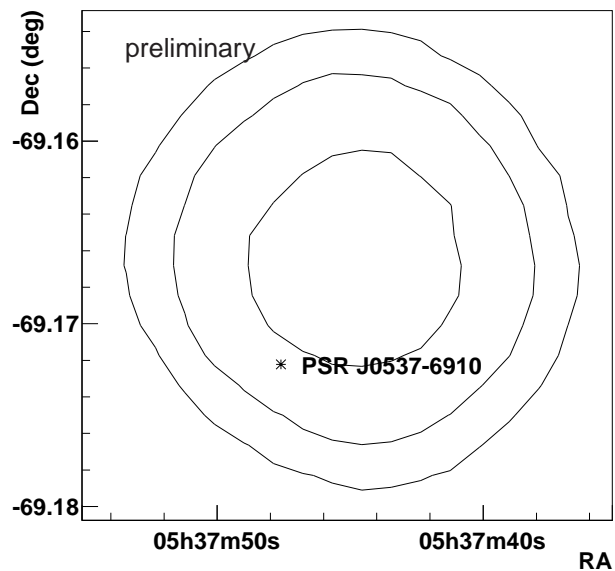}
\put(20,85){\bf \sf preliminary}
\end{overpic}

  \caption{Best fit position of the VHE gamma-ray source in the LMC. The contours denote 68\%, 95\% and 99\% confidence level of the position of a point-like source.
  }
  \label{fig:position}
 \end{figure}

A model of a point-like source has been fit to the sky map in order to obtain the best-fit position of the source at RA $5^{\mathrm{h}}37^{\mathrm{m}}(44 \pm 2)^{\mathrm{s}}$, Dec $69^\circ9'(57 \pm 11)''$ (epoch J2000). Figure~\ref{fig:position} shows the confidence contours of the fit-position. The position of the pulsar PSR\,J0537$-$6910 is inside the 95\% contour. X-ray observations of this object show \cite{Chen} non-thermal emission from the pulsar wind nebula around PSR\,J0537$-$6910 and only thermal emission from the supernova ejecta. It seems justified to attribute the gamma-ray emission entirely to the PWN and to interpret it as Inverse Compton emission from the nebula powered by the pulsar. This is the first, and so far only, extra-Galactic PWN seen in VHE gamma rays.

\begin{figure}[!t]
  \vspace{5mm}
  \centering
\begin{overpic}[width=0.5\textwidth]{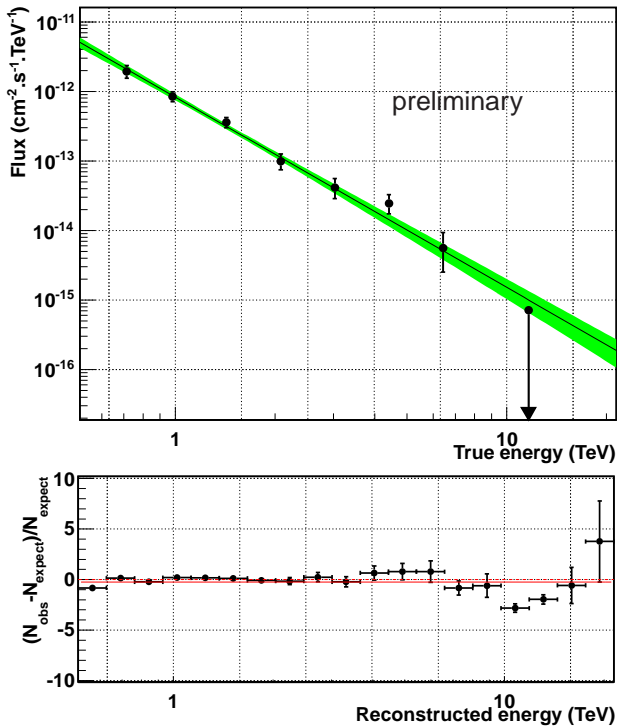}
\put(50,80){\bf \sf preliminary}
\end{overpic}  \caption{Energy spectrum of the pulsar wind nebula N\,157B. The top panel shows the best fit model and data points in true energy, the bottom panel represents the residuals in reconstructed energy.}
  \label{fig:spectrum}
 \end{figure}

Figure~\ref{fig:spectrum} shows the energy spectrum at TeV energies of N\,157B. It follows a power law
\begin{equation}
\frac{dN}{dE} = \Phi_{1\,\mathrm{TeV}} \left( \frac{E}{1\,\mathrm{TeV}} \right) ^{-\Gamma}
\end{equation}
with a spectral index of $\Gamma = 2.7 \pm 0.2_\mathrm{stat} \pm 0.3_\mathrm{syst}$ and a flux normalisation at 1\,TeV of $\Phi_{1\,\mathrm{TeV}} = (8.3 \pm 0.8_\mathrm{stat} \pm 2.5_\mathrm{syst}) \, 10^{-13} \,\mathrm{TeV}^{-1}\mathrm{cm}^{-2}\mathrm{s}^{-1}$. The energy flux in the 1--10\,TeV energy range is $F_{1-10\,\mathrm{TeV}} = 1.5 \, 10^{-12} \,\mathrm{erg\,cm}^{-2}\mathrm{s}^{-1}$. 
The apparent efficiency of converting pulsar spin-down luminosity to TeV gamma rays is then $\epsilon_{1-10\,\mathrm{TeV}} = 0.08 \dot{E}$. This is comparable to other pulsar wind nebulae found in the Milky Way \cite{PWN}. This particular PWN is detected at a distance of 48\,kpc thanks to its high spin-down luminosity.

\section{The Supernova Remnant SN\,1987A}

 \begin{figure}[!t]
  \vspace{5mm}
  \centering
\begin{overpic}[width=0.5\textwidth]{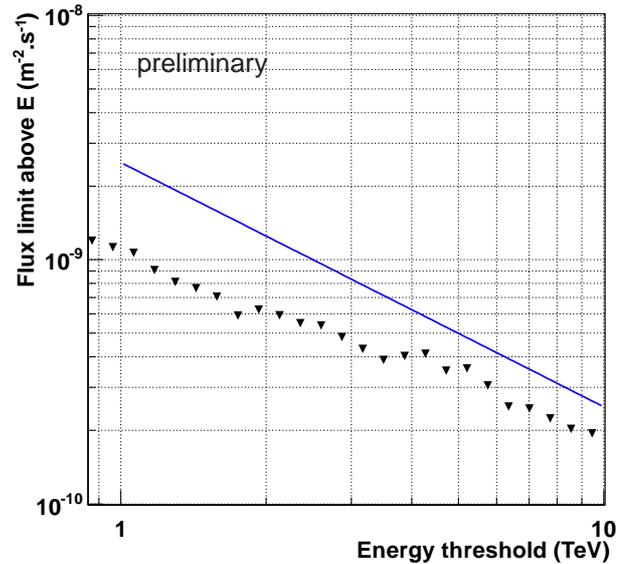}
\put(20,80){\bf \sf preliminary}
\end{overpic}
  \caption{Integrated upper limits on the VHE gamma-ray emission from SN\,1987A. The upper limits are obtained for a spectral index of 2 at 99\% confidence level. The blue line denotes the prediction for gamma-ray emission from hadronic interactions at present time for a variable flux increasing with time \cite{BKV2011}.
  }
  \label{fig:UL}
 \end{figure}

The supernova remnant SN\,1987A is located only $0.22^{\circ}$ away from N\,157B, making a separate analysis of this object difficult. A point-like source analysis at the position of SN\,1987A reveals 40 excess events with a statistical significance of $3.8\,\sigma$. A spill-over of events from the nearby source cannot be excluded. An upper limit on the integrated flux assuming a spectral index of 2 can be derived to be $F(>1\mathrm{TeV}) < 1.1\,10^{-13} \,\mathrm{cm}^{-2}\mathrm{s}^{-1}$ (99\% confidence level), assuming a constant flux. Figure~\ref{fig:UL} shows the upper limit for different energy thresholds.

The gamma-ray flux above 1\,TeV originating from hadronic interactions in SN\,1987A is predicted to be around $2.5\,10^{-13} \,\mathrm{cm}^{-2}\mathrm{s}^{-1}$ at the present time \cite{BKV2011}. Constant emission at this level could be excluded with the current data set. However, the data was mainly accumulated over the last 6 years during which the emission is expected to rise. Therefore, a time-dependent analysis will be performed. Further on, the expected flux increase may lead to a detection in the near future.

%\begin{table}[t]
%\begin{center}
%\begin{tabular}{l|ccc}
%\hline
%\hline
%year & differential flux $> 3$\,TeV  \\
%	 & $10^{-13}$ [$\mathrm{cm}^{-2}\mathrm{s}^{-1}$] \\
%\hline
%2005 & $< 1.3$ \\
%2007 & $< 0.74$ \\
%2008 & $< 0.79$ \\
%2009 & $< 0.67$ \\
%2010 & $< 0.78$ \\
%2011 & $< 1.1$ \\
%\hline
%all & $< 0.48$ \\
%\hline
%\hline
%\end{tabular}
%\caption{Upper limit (99\% confidence level on the flux from the supernova remnant SN\,1987A. A threshold energy of 3\,TeV was chosen for better comparison with the predictions of \cite{BKV2011}. }
%\label{tab:UL}
%\end{center}
%\end{table}

\section{Large-scale Diffuse Emission}

The \textit{Fermi}/LAT telescope discovered MeV and GeV gamma-ray emission from the direction of the LMC which is found to be correlated with  the massive star forming region 30 Doradus and interpreted as arising from Cosmic Ray interactions in this star forming region. No emission from point-like sources correlated with either N\,157B or SN\,1987A has been detected. The large extension of the \textit{Fermi} source of about $5^\circ \times 5^\circ$ does not allow any conclusion on a possible TeV emission with the current HESS data set. More sophisticated observation procedures, like a mini-scan of the LMC, would be needed to observe this object.

\section{Conclusion}

The LMC is observed with H.E.S.S. on an annual basis since 2003. In this data the first extra-galactic TeV pulsar wind nebula has been detected: N\,157B. The powering pulsar PSR\,J0537$-$6910 is the most energetic pulsar known, its apparent efficiency of converting spin-down luminosity to TeV gamma-rays is $0.08 \dot{E}$. It is thus comparable with Galactic PWNe. The young supernova remnant SN\,1987A is not detected in the current data set. The derived upper limit exclude constant emission at a level predicted for the current epoch; however, the expected rise of the emission requires a time-dependent analysis. The current HESS data set does not allow to test for large-scale diffuse emission from 30 Doradus similar to what was found by \textit{Fermi}.

The LMC remains a very interesting target for TeV gamma-ray emission with current and future instruments. HESS-II, the extension of HESS which will start operation in 2012, will reduce the energy threshold and increase the sensitivity. Observing N\,157B at GeV energies will improve the understanding of the emission of this object. With a better sensitivity it may be possible to finally detect gamma-ray emission from SN\,1987A. This supernova remnant is the prime candidate for testing when Cosmic Ray acceleration in supernova remnants starts. The future Cherenkov Telescope Array (CTA) will bring an improved sensitivity with a better angular resolution which may help to separate the emissions from  N\,157B and SN\,1987A. 

\section*{Acknowledgements}
The support of the Namibian authorities and of the University of Namibia
in facilitating the construction and operation of H.E.S.S. is gratefully
acknowledged, as is the support by the German Ministry for Education and
Research (BMBF), the Max Planck Society, the French Ministry for Research,
the CNRS-IN2P3 and the Astroparticle Interdisciplinary Programme of the
CNRS, the U.K. Science and Technology Facilities Council (STFC),
the IPNP of the Charles University, the Polish Ministry of Science and 
Higher Education, the South African Department of
Science and Technology and National Research Foundation, and by the
University of Namibia. We appreciate the excellent work of the technical
support staff in Berlin, Durham, Hamburg, Heidelberg, Palaiseau, Paris,
Saclay, and in Namibia in the construction and operation of the
equipment.

%\vspace{\baselineskip}

\clearpage

\end{document}